# Using Conformity to Probe Interaction Challenges in XR Collaboration

**Jeremy Hartmann**[*]
University of Waterloo
Waterloo, ON N2L 3G1, Canada
j3hartma@uwaterloo.ca

**Hemant Bhaskar Surale**[*]
University of Waterloo
Waterloo, ON N2L 3G1, Canada
hsurale@uwaterloo.ca

**Aakar Gupta**
University of Waterloo
Waterloo, ON N2L 3G1, Canada
aakar.gupta@uwaterloo.ca

**Daniel Vogel**
University of Waterloo
Waterloo, ON N2L 3G1, Canada
dvogel@uwaterloo.ca

## ABSTRACT
The concept of a conformity spectrum is introduced to describe the degree to which virtualization adheres to real world physical characteristics surrounding the user. This is then used to examine interaction challenges when collaborating across different levels of virtuality and conformity.

## INTRODUCTION
Milgram et al.'s reality-virtuality continuum [2] provides a means to classify and compare different types of virtualization technologies in XR (i.e. AR, VR, MR). We propose an orthogonal axis to this continuum, one that incorporates the degree to which the virtualization adheres to real world physical characteristics surrounding the user. We call this axis the *Conformity Spectrum*. Our motivation stems from an apparent lack of nomenclature to describe certain XR environments; such as when the geometric representation of virtual objects are directly mapped to real physical objects occupying the same space. In such a scenario, the virtual world conforms to immediate physical reality creating a virtual-conforming reality. We believe this perspective will foster discussion and uncover areas for research. In this paper, we use conformity to examine interaction challenges when collaborating across XR.

## CONFORMITY SPECTRUM
The conformity spectrum captures how closely interaction entities [1] in the user's locus of attention adhere to real world physical characteristics. In this paper, we focus on conformity to time, space, and geometry, but other characteristics like material texture, tactile texture, mass, sound, and smell could be included as well (assuming virtualizing technologies can synthesize these dimensions).

[*]Authors have contributed equally to this work.

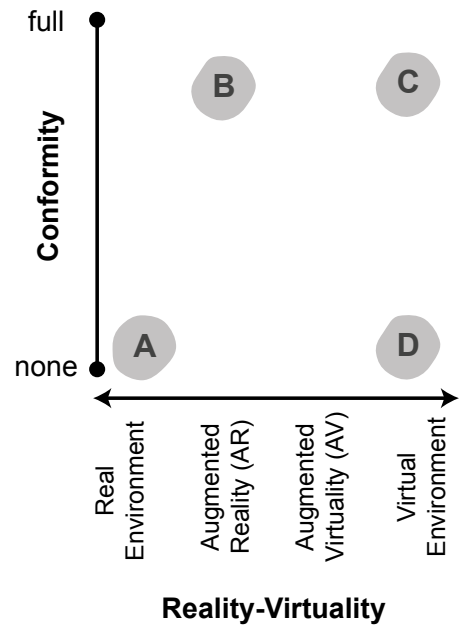

Figure 1. Conformity spectrum as an orthogonal axis to Milgram et al.'s reality-virtuality continuum. In the accompanying text, we consider four usage contexts in this space plotted as grey areas: (A) using a tablet in reality as an example of no conformity; (B) AR where some objects are virtual, but most are physical to capture high conformity; (C) and VR where virtual objects are perfectly aligned with real objects to demonstrate high conformity; (D) VR with no physical objects as an exemplar of no conformity.

Figure 1 illustrates the degree of conformity as a vertical axis orthogonal to Milgram et al.'s reality-virtuality continuum. The corners represent interesting extremes of this space. In the bottom-left are digital interactions that have no connection to the immediate physical environment, e.g. using conventional computers. The upper-left represents perfect Spatial Augmented Reality (SAR) where all digital interactions have a one-to-one mapping with physical geometry. The upper-right represents a fully virtual environment perfectly conforming to the immediate real environment. Finally, the bottom-right is what most consider current VR, a fully virtual environment that does not conform to the immediate real environment in any way.

To further explain conformity, consider four usage contexts (labelled areas in Figure 1 and illustrated in Figures 2 and 3). Each consists of a simple environment where a user interacts with four blocks sitting on a table. In *Context A*, the user interacts with all blocks using a standard tablet interface (e.g CAD program), but real blocks are also nearby. The user is in reality, but there is no conformity between their interaction and the real environment. In *Context B*, the user interacts through a see-through AR display with two virtual blocks beside two real blocks, creating a high conformity interaction. In *Context C*, the user interacts through a VR Head Mounted Display (HMD) where two virtual blocks and the table perfectly conform to their real counter-parts in the room (approaches like [4] can be used to generate such environments). In *Context D*, the user interacts through a VR HMD and all blocks (and the table) are entirely virtual, but the real corresponding blocks and table might be elsewhere in the room. This is the classic VR usage context (e.g. [3]) with no conformity.

**FACILITATING COLLABORATION**

We use the conformity spectrum as a lens to investigate remote collaboration between these usage contexts. Since each user experiences the same task environment differently, this adds to the complexity of designing an optimal collaborative interface. The conformity spectrum provides a structured way to identify when collaboration breaks down and identify techniques to mitigate breakdowns.

For exploration, we employ an abstract task where two remote users collaboratively manipulate four blocks. The degree to which the blocks conform to the real world differs depending on where the usage context lies in the conformity spectrum. The collaborators need to coordinate the movement of blocks into some final configuration, and both users have the potential to manipulate any, or all blocks. This abstract task represents real world collaborations centered around the need for both parties to have a shared experience with a physical artifact. Specific examples include furniture assembly, home repair, and machine operation. The ultimate goal is to transform the real world artifact into a goal state, such as a fully assembled bed, a repaired kitchen sink, or successfully operating a 3D printer.

Fundamentally, remote collaboration relies on two aspects: 1) the amount of *awareness* of each others actions to communicate movements and prevent simultaneous access to the same resource; and 2) the quality of *synchronization* of the shared state of the task across both collaborators. In addition to these general task requirements, we hypothesize that the collaborative user experience is a function of the *relative conformity* between different usage contexts and objects within those environments. We use awareness and synchronization as metrics to examine collaboration under different levels of relative conformity.

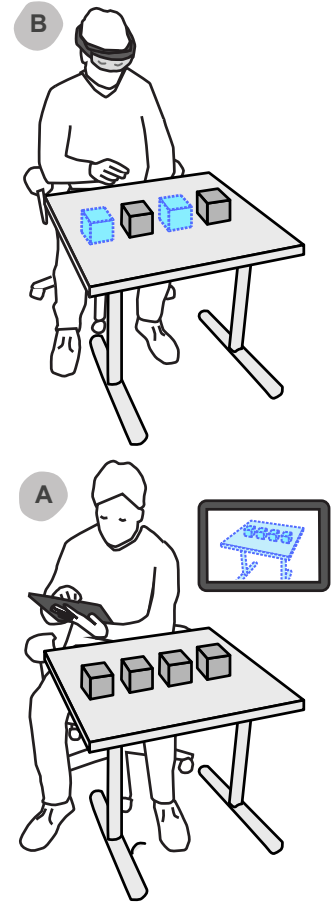

**Figure 2.** *Context A*, the user interacts with blocks using a standard tablet interface while real blocks are nearby; *Context B*, the user interacts in AR where they see two virtual blocks and two real blocks on a real table.

**Scenarios**

In the following scenarios, collaborators are refereed to as **A**, **B**, **C**, or **D** according to usage context (see Figures 2 and 3). Each task block can be in one of four states from a single collaborator's perspective: 1) *owned*, the block conforms to the real world, but is a virtual block for the remote collaborator; 2) *shared*, the block conforms for both collaborators; 3) *unowned*, the block is purely virtual and conforms for the remote collaborator; and 4) *orphaned*, where the block is purely virtual for both collaborators. Table 1 enumerates how each scenario relates to each block state.

*Scenario 1: A ⟺ B*

**A** uses a tablet with low conformity and **B** is in AR with high conformity. If either user moves an orphaned block, there are no synchronization or awareness issues. If **A** moves an unowned block using the tablet, **B**'s AR environment becomes out-of-sync since the block was previously conforming to their real environment. For **B**, the real block and its now virtual moved version diverge, making awareness conflicting. However, if **B** moves an owned block, there will be no awareness or synchronization problems for **A**. To address the awareness issue, the block displacement for **B** must be clearly visualized (e.g. colour change or ghosting). For synchronization, **B** must be prompted to physically move the physical block to the new virtual block position so conformity is reestablished. Note that due to the low conformity usage context for **A**, they will have to manually re-sync all their real blocks to ultimately complete the real task. Essentially, low conformity means synchronization is postponed until after collaboration.

|   | Orphaned | | Owned | | Unowned | | Shared | |
|---|---|---|---|---|---|---|---|---|
|   | *Aware* | *Sync* | *Aware* | *Sync* | *Aware* | *Sync* | *Aware* | *Sync* |
| A | ● | ● | - | - | ● | ● | - | - |
| B | ● | ● | ◐ | ○ | ● | ● | ◐ | ○ |
| C | ● | ● | ● | ○ | ● | ● | ● | ○ |
| D | ● | ● | - | - | ● | ● | - | - |

Table 1. **Effect of awareness and synchronization when manipulating blocks in different conformity states (● good, ◐ partial, ○ poor, - not applicable.)**

*Scenario 2: A ⟺ C*
**A** is collaborating with **C**, who is in VR with high conformity. Again, for **A**, block states can only be unowned or orphaned. Like Scenario 1, if either user moves an orphaned block, or **C** moves an owned block, there are no immediate synchronization or awareness issues. However, when **A** moves an unowned block using their tablet, the effect on **C** is more nuanced that Scenario 1. For **C**, awareness remains clear since the once conforming block is now virtual. However, synchronization is even more challenging since **C** may not know the block no longer conforms until they manipulate it. To address the synchronization issue, feedback must provide visual cues to indicate a block has been moved and no longer conforms.

*Scenario 3: C ⟺ D*
Both users are in VR, but only **C** has high conformity. If **C** is to move an owned block, they simply reach into the world and move it. In so doing, **D**'s VE will immediately update as the virtual space between the two collaborators is shared. If **D** moves a block, the virtual space is updated instantaneously but **C**'s physical environment desynchronizes from their shared state. In such a case, **D**'s action could almost be thought as a suggestion to **C**, as the agency on the physical proxies is only within **C**'s domain.

*Scenario 4: B ⟺ C*
**B** is in AR and **C** is in VR, both have high conformity. In this scenario, we look at three situations. First, if **B** moves an owned block, **C**'s virtual environment can be immediately updated maintaining their shared state. Second, if **B** moves a shared block, **B**'s state will remain synchronized but **C**'s environment will become desynchronized. In such case, **C** must re-sync their physical environment to reestablish conformity. Third, if **B** moves an unowned block, then **C** has the same problem as with the shared block. To help prevent desynchronization, a constraint on the task could be that each collaborator can only move owned blocks, in which case the shared state will always be in sync; but if shared or unowned blocks are moved, then some mechanism needs to be put in place to maintain shared state.

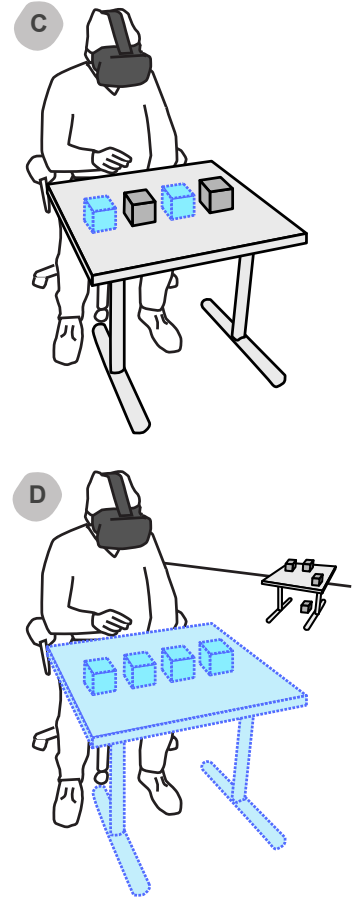

Figure 3. *Context C*, the user interacts in VR with two virtual blocks and a table that perfectly conform to their real counter-parts, and two purely virtual blocks; *Context D*, the user interacts in VR where everything is entirely virtual, but real blocks and table are nearby.

*Scenario 5: B ⟺ D*
**B** is in AR with high conformity and **D** is in VR with no conformity. If **B** moves an owned or orphaned block, the shared state remains synchronized and **D** is immediately notified of the change. If **D** moves a block, **B** can be immediately notified of the change but the shared state between the two collaborator would become out of sync. This type of interaction raises some interesting issues, as the interaction of **D** has no real effect on the state of **B**'s physical proxies, at best they are just a suggestion from **D**. The agency of action remains in **B**'s control and they can ultimately decide whether to integrate **D**'s changes or not.

*Scenario 6: A ⟺ D*
Since both collaborators have no conformity, there is no agency on the blocks in the 'real' world since all blocks are in the orphaned state. Awareness and synchronization of the shared virtual environments is trivial. However, since the task is to move real blocks into an agreed upon form, once both parties accept the current shared virtual state, they need to exit the interaction and synchronize the physical blocks irrespective of each other.

## CONCLUSION AND FUTURE WORK

The scenarios in this paper describe a single task and the ways in which conformity can help design the interaction space. However, the *Conformity Spectrum* is not limited to a single task, many other interactions with different constraints can be studied using this style of analysis. Future work will investigate and evaluate different tasks defined on this continuum, and we are preparing a series of usability experiments to explore remote and co-located collaboration under different levels of conformity.